*A slightly modified version of this report is in the process of submission and evaluation in a peer-reviewed journal*

# Mathematical model for autoregulated miRNA biogenesis

Dimpal A Nyayanit[1,2] and Chetan J Gadgil[1,2,3,*]

[1]Chemical Engineering and Process Development Division, CSIR-National Chemical Laboratory, Dr. Homi Bhabha Road, Pune 411008

[2]Academy of Scientific and Innovative Research, CSIR-NCL, Pune 411008

[3]CSIR-Institute of Genomics and Integrative Biology, Mathura Road, New Delhi 110020

*Address for communication

Chemical Engineering Division, National Chemical Laboratory,

Dr Homi Bhabha Road, Pune 411008

Maharashtra, India

Phone: 91-20-25902163

Fax: 91-20-25902621

Email: cj.gadgil@ncl.res.in

**Running title**

Mathematical model for miRNA biogenesis

**Abstract:**

MicroRNAs are small non-coding nucleotide sequences that regulate target protein expression at post-transcriptional levels. Biogenesis of microRNA is a highly regulated multi-step pathway. Regulation of miRNA biogenesis can be caused directly by the components of the biogenesis pathway or indirectly by other regulators. In this study, we have built a detailed mathematical model of microRNA biogenesis to investigate the regulatory role of biogenesis pathway components. We extended a previous model to incorporate Microprocessor regulation of DGCR8 synthesis, exportin-mediated transport to the cytoplasm, and positive auto-regulation catalysed by mature miRNA translocation into the nucleus. Our simulation results lead to three hypotheses (i) Biogenesis is robust to Dicer protein levels at higher Exportin protein levels; (ii) Higher miRNA transcript formation may lead to lower RISC levels: an optimal level of both precursor miRNA and Dicer is required for optimal miRNA formation at lower levels of Exportin protein; and (iii) The positive auto-regulation by mature miRNA translocation into the nucleus can decrease the net functional cytoplasmic miRNA. Wherever possible, we compare these results to experimental observations not used in the model construction or calibration.

**Statement of significance**

MicroRNA biogenesis involves competition for export of the precursor RNA, and autoregulation by mature miRNA translocation to the nucleus. We extend previous mathematical models and formulate a comprehensive model for miRNA biogenesis that includes these processes. Simulations reveal non-obvious scenarios, including the possibility that increase in transcript formation may not necessarily result in an increase in mature miRNA. Model simulations also result in testable predictions of the robustness to change in Dicer protein levels, and the existence of an optimal level of precursor and processor RNA for maximizing steady state mature miRNA concentrations.

## Introduction

MicroRNAs are small (~22 nucleotides) endogenous RNA that can bind to 3' UTR of their target mRNA causing mRNA repression or degradation (1, 2). They are part of a complex regulatory network and play diverse roles in controlling development (3), differentiation (4), and disease (5, 6). Micro-RNAs are themselves synthesized through a multi-step process which at least for some microRNA is an auto-regulated process (7).

MicroRNA (miRNA) biogenesis (Figure 1) begins with transcription by RNA polymerase II or RNA polymerase III (8, 9) to form a long transcript called primary miRNA (pri-miRNA). This transcript is then catalytically processed to precursor miRNA (pre-miRNA) by Microprocessor complex (MP), a protein heteromer consisting of proteins Drosha and DiGeorge Syndrome Critical Region 8 (DGCR8) (10, 11). Pre-miRNA is exported to the cytoplasm, with the help of transport protein Exportin 5 (Xpo5) (12).

In the cytoplasm, pre-miRNA is further acted upon by Dicer, which cleaves it to generate an approximately 20-22 nucleotides short duplex. The duplex comprises of a guide strand, which is complementary to the target mRNA, and the passenger strand which is eventually eliminated. This duplex, through association with a multiprotein complex including Argonaute (Ago), forms a ribonucleoprotein complex called the RNA-induced silencing complex (RISC). A member of the Argonaute protein family (e.g. AGO2 in humans) is the core of this RISC. The minimal functional RISC has been shown to comprise of Ago protein bound to a small RNA (13). The RISC bound duplex undergoes multiple steps which lead to degradation of passenger strand and results in a functional RISC (also called mature RISC or simply RISC) that can mediate miRNA effect on target mRNA. In this paper, we limit our analysis of miRNA biogenesis till binding of the duplex to Ago protein. We assume that this bound duplex is equivalent to mature RISC, referred to henceforth as RISC.

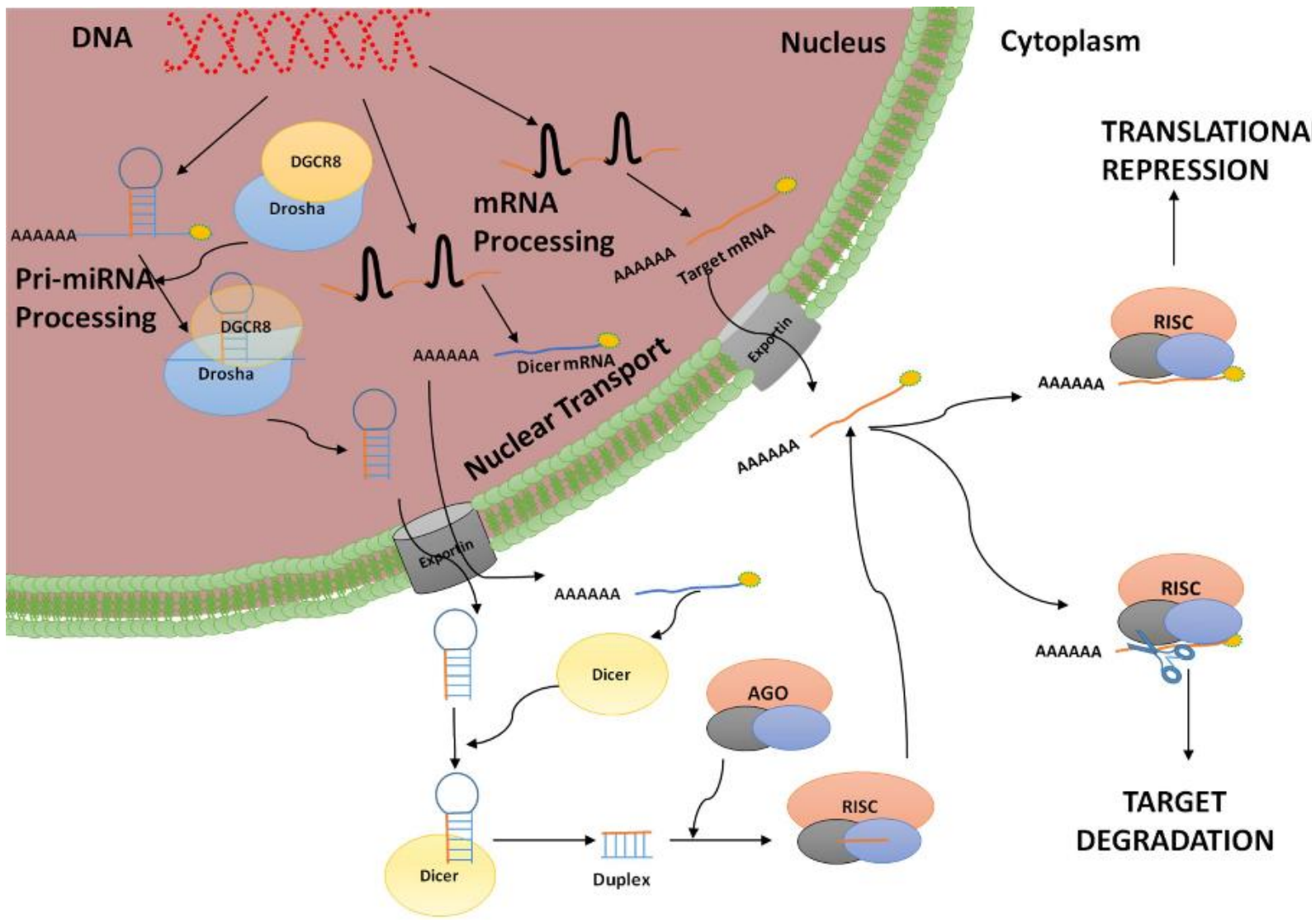

***Figure 1**: **Biogenesis of microRNA**. Transcription by RNA polymerase give rise to primary miRNA which is further processed to form precursor miRNA with the help of MP in the nucleus. Precursor miRNA formed is exported to the cytoplasm and following a series of step, along with the help of different enzymes, mature miRNA is generated.*

Biogenesis of miRNA is a regulated process. Regulation of the biogenesis pathway is achieved by regulating the amount, activity and location of pathway components such as proteins catalyzing the various biogenesis steps. A comprehensive overview of the regulators for the biogenesis pathway can be found in an excellent review by Siomi (14). Regulation of these pathway components can be through external elements such as transcription factors, or through auto-regulation either by the same component or by other pathway components. Here we focus on auto-regulation of miRNA biogenesis. MP levels are controlled through regulation of its constituents DGCR8 and Drosha. MP negatively regulates transcription of *dgcr8*-mRNA, while DGCR8 inhibits the degradation of Drosha protein (15). Similarly, *dicer*-mRNA and all pre-miRNA in the nucleus are thought to compete with each other for Exportin protein binding thereby regulating each other (16). Increase in Argonaute protein levels results in increased stability of miRNA by increasing their half-life (17). The Pasquinelli group has demonstrated positive auto-regulation of let-7 miRNA, where mature RISC, through binding to the primary transcript promotes the processing of its own pri-miRNA (7). It has been shown that Importin 8 (Imp8) regulates the transport of RISC to the nucleus (18). Nonlinear phenomena such as oscillations in miRNA levels have been experimentally observed, though these may be due to sequence-specific regulation rather than a mechanism common to miRNA biogenesis (19). Nevertheless, understanding the contributions of each of the regulatory modes to miRNA biogenesis will require a framework to simultaneously assess their effects, which is provided by a mathematical model of microRNA biogenesis.

Several mathematical models have been used to study the interaction between single miRNA-single mRNA, single miRNA-multiple mRNA, multiple miRNA-single mRNA, and multiple miRNA-multiple mRNA (20–33). However, to date, there are only a few models for miRNA biogenesis. Barad and co-workers (34) studied the efficacy and specificity of transcript cleavage by MP. They modeled *dgcr8*-mRNA formation rate as being inhibited by MP levels, and MP formation rate as being dependent on DGCR8 and Drosha levels. Through a steady-state analysis, they concluded that MP levels adjust to pri-miRNA levels through auto-regulatory feedback ensuring that unbound microprocessors are kept in the optimal range (34). Wang and co-workers (35) built a comprehensive chemical reaction representation of unregulated miRNA biogenesis and carried out deterministic and stochastic simulations to show that the miRNA noise is similar for both the mechanisms of repression (mRNA cleavage and inhibition of translation). Through a sensitivity analysis, they identified miRNA and mRNA transcription rate and RISC decay as key parameters affecting many reaction rates in the pathways (35). Since the focus of the Barad model is MP formation from DGCR8 and Drosha and its activity in cleaving RNA including *dgcr8*-mRNA, it does not include details of miRNA biogenesis. The Wang model is a more detailed model that includes several aspects of miRNA biogenesis and activity, but it does not include any auto-regulation since several of these autoregulatory mechanisms were identified after the publication of that model.

Here we seek to build a mathematical model to study the effect of regulation on miRNA biogenesis dynamics. The miRNA biogenesis model of Wang is extended to include details of MP regulation as experimentally shown by Han et al.(15), and regulation due to competition between *dicer*-mRNA and pre-miRNA for exportin binding as shown by Bennasser et al.(16). We also extended the model through the incorporation of Imp8-assisted nuclear transport of mature RISC and its effect on miRNA processing to include the experimentally reported (7) positive auto-regulation of microRNA. We simulate the model with varying levels of Dicer, Exportin and levels of regulation. Our simulation results suggest that levels of mature RISC are robust to Dicer protein levels when Exportin levels are high. At lower Exportin levels, an optimal level of Dicer and pre-miRNA exists, and the concentration of mature RISC decreases with any change from these optimal levels, implying that the mature RISC level may decrease even if the primary miRNA transcription increases. We also find that increasing the strength of the positive auto-regulation of miRNA biogenesis by mature RISC may result in lower or higher mature RISC levels depending on the relative rate of translocation to the nucleus and the enhancement of pri-miRNA processing by MP bound to this translocated miRNA. Using additional simulations we propose mechanism-based explanations for these nonintuitive observed effects.

**Results:**

**A model for regulated miRNA biogenesis:**

Through the incorporation of additional biochemical details, the effect of regulation on miRNA biogenesis is included in the Wang model for miRNA biogenesis, shown in Figure 2. A comparative analysis of our reaction system with the Wang and Barad models is given in Table 1. Formation of pri-miRNA (Pri), *dgcr8* mRNA (mDGCR8), *drosha*-mRNA (mDrosha), Xpo5,

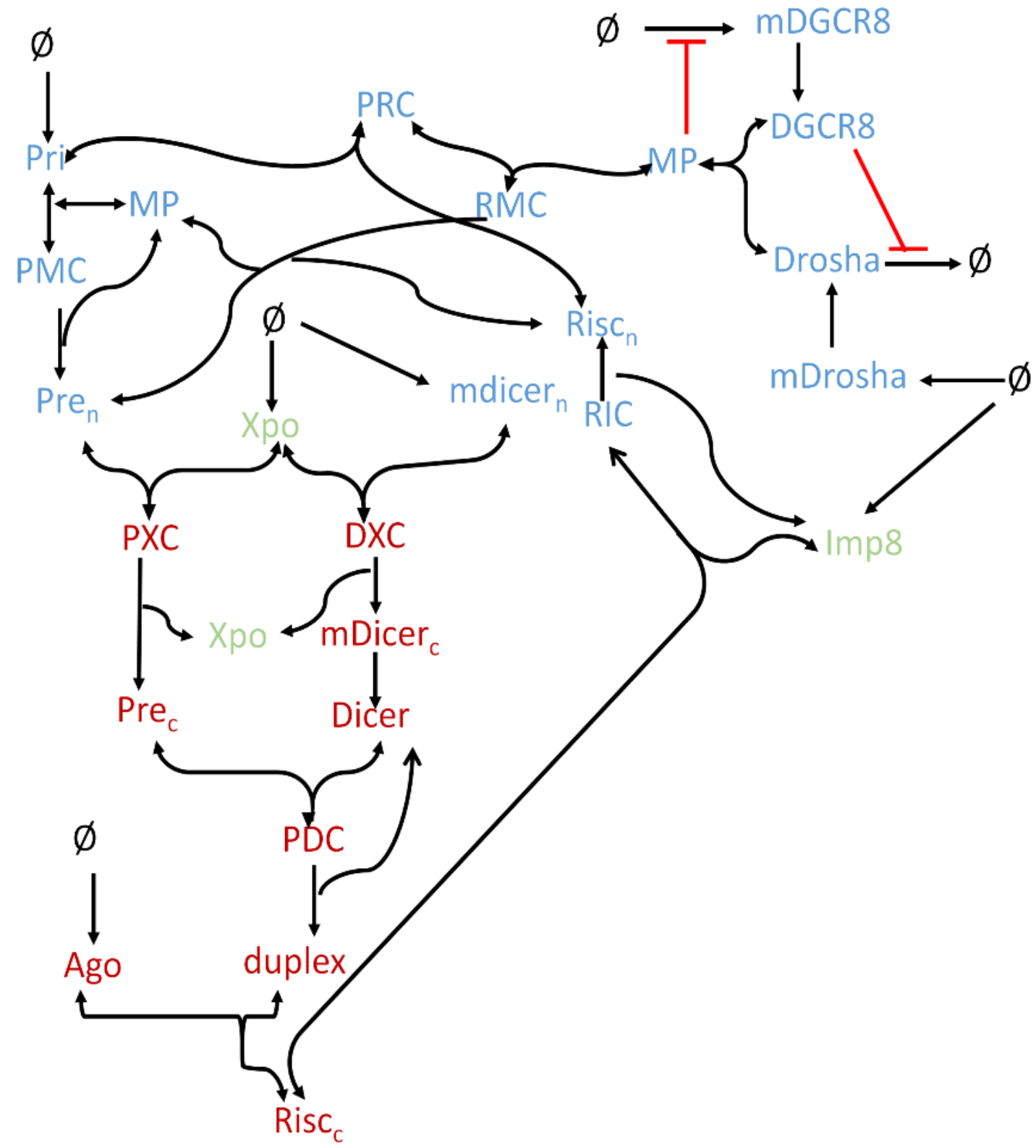

***Figure 2: Schematic representation of regulated model of miRNA biogenesis****. RISC$_C$ is the mature RISC. Red lines (⊥) indicate negative regulation. Species in blue and red colour are present in nucleus and cytoplasm respectively. Exportin and Importin molecules are assumed to be at the nuclear membrane pore. ∅ denotes Source or Sink.*

Argonaute (Ago) and Importin8 (Imp8) is modeled as a zero order birth process with rates $b_{pri}$, $b_{dgm}, b_{drm}, b_{xpo5}, b_{ago}, b_{\mathrm{Im}p8}$ respectively. Mass action kinetics is used for the association, dissociation, and degradation reactions. The pri-miRNA to pre-miRNA processing by MP is assumed to follow a Michaelis-Menten kinetics mechanism with MP as the enzyme. Reversible association of pri-miRNA and MP forms a pri-MP complex (PMC), which is followed by irreversible formation of pre-miRNA and recycling of MP. Similarly, transport of pre-miRNA and *dicer*-mRNA to the cytoplasm is assumed to be 'catalyzed' by Exportin. In the cytoplasm, *dicer*-mRNA is translated to Dicer protein, which enzymatically catalyzes the conversion of pre-miRNA to Duplex. Duplex reversibly associates with Argonaute leading to the formation of RISC (mature miRNA). For simplicity, Argonaute is considered to be representative of all the proteins required for the activity associated with RISC. *DGCR8*-mRNA and *Drosha*-mRNA are translated to form DGCR8 and Drosha respectively. Both of these protein associate forming MP protein that catalyzes the pre-miRNA formation. For simplicity, we do not model transport of DGCR8 and Drosha transcripts, even though they may also compete for Exportin protein; as well as transport of these proteins to the nucleus.

The model includes regulation of MP formation, in which DGCR8 protein inhibits Drosha degradation and MP leads to repression of *Dgcr8* transcription. Positive regulation by RISC is modeled by including the transport of RISC to the nucleus catalyzed by Importin (Imp8)[17]. Inside the nucleus, RISC binds to pri-miRNA forming a pri-risc complex (PRC). PRC associates with MP forming pri-risc-mp complex (RMC) which produces pre-miRNA with higher efficiency than the pri-MP complex (PMC) ($b^{n}_{prn} > b^{n}_{pre}$).

Table 2 lists the differential equations arising from a mass balance for each of the constituents of the positively regulated miRNA biogenesis system. The parameter values for numerical simulation listed in Table 3 are mostly taken from Shimoni et al (36), or related models, or assumed based on the reported values for similar processes. Our conclusions from this study are independent of the absolute values of the parameters used for simulation. As the system contains 24 equations and is non-linear, analytical solution for the dynamics or the steady-state concentrations could not be found. We have performed numerical simulations using MATLAB (MathWorks, version R2012a) and symbolic calculation using Mathematica (Wolfram, version 7).

**RISC levels are robust to changes in Dicer levels at high Exportin protein levels.**

Pre-miRNA and *dicer*-mRNA present in the nucleus compete with each other for transport protein (Xpo5) binding, which exports them to the cytoplasm. We numerically studied the effect of this competition on RISC levels, for various combinations of miRNA, Dicer and Exportin levels. Figure 3A shows the effect of increasing $b_{dic}^{n}$ (different color lines) with varying pri-miRNA formation rate $b_{pri}^{n}$ (X-axis) at different Exportin formation rates (different subplot rows). We varied the rate of pri-miRNA formation at several levels (5-fold lower and higher than the reference value indicated in Table 3) of *dicer*-mRNA formation rate ($b_{dic}^{n}$) to analyze the effect of pri-miRNA over-expression on Dicer as well as RISC steady state. We have also analyzed the system at different expression levels of Exportin. Each row in Figure 3A has the same Exportin formation rate, which increases from top to bottom rows. It is seen from the simulations that over-expression of pri-miRNA leads to decrease in the Dicer levels at lower levels of Exportin formation rate, as depicted in Figure 3A (top and middle rows, right column).

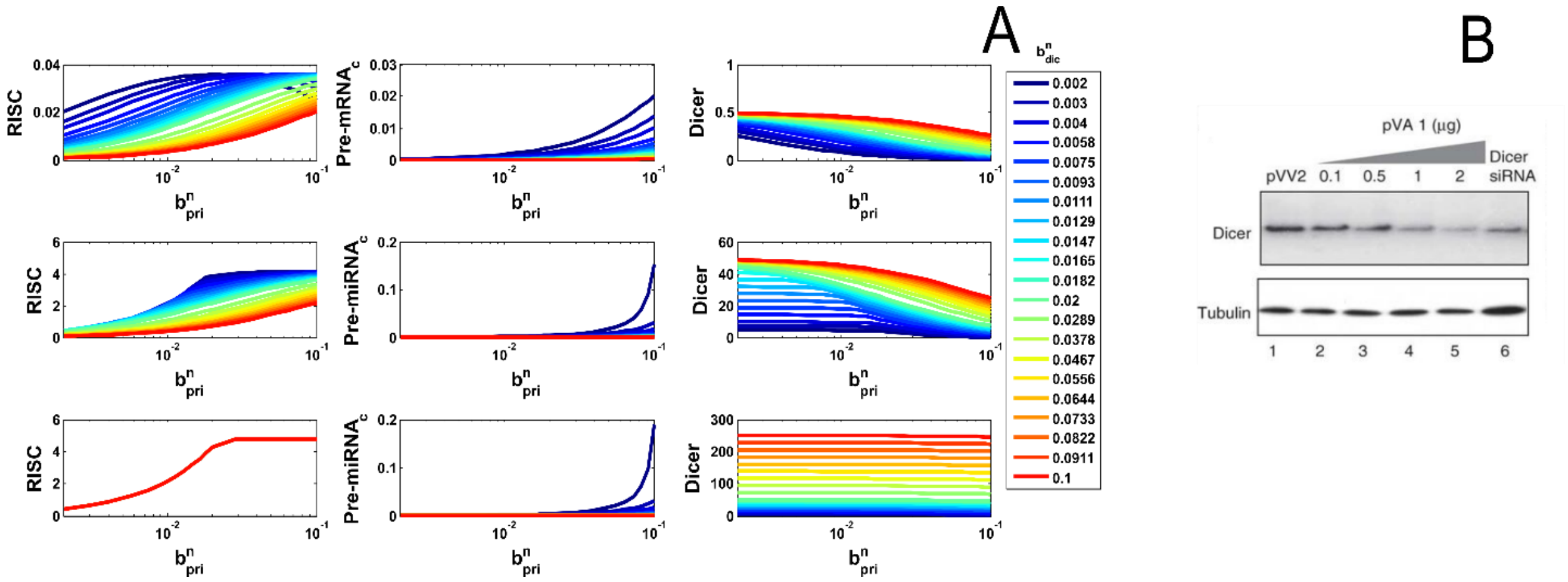

***Figure 3: Steady state levels of RISC, pri-miRNA and Dicer at varying levels of pri-miRNA ($b_{pri}$) and dicer-mRNA ($b_{dic}$) formation.*** *Varying colored lines are different $b_{dic}$ . Rows from top represent increasing levels of Exportin ($b_{xpo5}$) . Middle row has Exportin formation rate $b_{xpo5}$ = 0.01 molecules/s which is the reference value indicated in Table 3. Top row is 100 times lesser and bottom row is 100 times higher than the reference value indicated in Table 3 for the $b_{xpo5}$ value. *denotes value corresponding to parameter set given in Table 3. B) Reprint of Figure 4b is from Bennasser et al 2011 (licence agreement pending) representing effect of varied small RNA concentration on dicer expression levels.*

When the Exportin formation rate (and consequently Exportin level) is high, Dicer levels remain unaffected by a change in pri-miRNA formation, as shown in Figure 3A (bottom row, right column).

In an experimental study (16) not used in the choice of parameters for this mathematical model, it was shown that over-expression of pri-miRNA leads to reduced levels of Dicer, as a result of Exportin saturation. Our simulation results qualitatively agree with the experimental observations (reproduced here as Fig 3B). The left column of Figure 3A shows the steady state level of RISC when these three factors are changed. For most cases, the RISC level as expected monotonically increases with the pri-miRNA formation rate. At low and intermediate levels of Exportin, for a given pri-miRNA formation rate, the RISC concentration is lower when the *dicer*-mRNA formation rate is high (Figure 3A, top and middle rows, left column). At high levels of Exportin, mature RISC steady state levels are robust to changes in the *dicer*-mRNA formation rate, as seen in Figure 3A (bottom row), where the overlapping colour graphs indicate that the response to increased pre-miRNA does not depend on the Dicer levels. Thus we predict that the results seen in the Bennasser et al. experimental study will not be observed at high Xpo5 expression levels. At these high Exportin levels, Dicer expression will be largely unaffected by pri-miRNA overexpression.

The use of a mechanism-based model enables us to investigate the cause of this observation. We carried out additional simulations to dissect the contributions of each factor. For limiting amounts of exportin, the amount of pre-miRNA and Dicer mRNA in the nucleus can limit the amount of the other transcript that can be transported. Increasing the formation of *dicer*-mRNA in the nucleus by increasing its formation rate leads to its accumulation within the nucleus leading to an increase in its level (data not shown) as compared to its competitor (pre-miRNA). This increased level of *dicer*-mRNA competes with the pre-miRNA in the nucleus for Exportin binding, which causes a larger quantity of the *dicer*-mRNA to be transported to the cytoplasm. This leads to higher Dicer-Exportin complex (DXC) concentration and hence unavailability of Exportin for transport of preMRNA. As a consequence, there is a decrease in the level of pre-miRNA-exportin complex (PXC) (Supplementary Figure 1A, 1B) and therefore lesser amount of pre-miRNA in the cytoplasm. Increase in the amount of Dicer protein and a decrease in the amount of pre-miRNA in the cytoplasm affects pre-dicer complex which decrease is also reflected downstream and can be observed in duplex (Supplementary Figure 1C) as well as RISC levels.

At higher levels of Exportin formation rate, increasing Dicer mRNA formation increases Dicer protein levels (Figure 3A bottom row right column), that enhances the association rate with pre-miRNA to form PDC. As PDC increases miRNA duplex levels also increase. It is also important to note that Dicer is a catalyst, and gets regenerated when PDC forms duplex. With the increase in Exportin and Dicer protein levels, RISC formation depends only on the pre-miRNA$_c$ level and RISC exhibits robustness to Dicer formation rate changes (left column bottom row).

Further, we predict (Supplementary Figure 2, which is the same as Fig 3 except for increased $b_{pri}^{n}$ range) that increasing the pri-miRNA formation rate $b_{pri}^{n}$will not lead to saturation of RISC levels (Supplementary Figure 2, left column top row). Rather, after the optimal value, increase in $b_{pri}^{n}$ will counterintuitively result in a decrease in functional RISC levels. This can be explained using simulations as resulting from competition for Xpo5. Higher pri-miRNA levels will result in higher nuclear pre-miRNA levels, which will compete with Dicer mRNA for export. This competition will lead to less Dicer mRNA being available for translation, and consequently less Dicer protein. The amount of Dicer will decrease as the competition grows with increasing $b^n_{pri}$. Therefore the amount of cytoplasmic pre-miRNA-Dicer complex will decrease, leading to a decrease in duplex and therefore RISC.

**Optimal levels of pre-miRNA affinity for Exportin is required for the maximum level of RISC**

In order to investigate other factors that that contribute to this unintuituve effect of decreased RISC resulting from increased $b^n_{pri}$, at different levels of Exportin formation rate, we numerically explored this effect by altering the binding affinity of pre-miRNA and *dicer*-mRNA for Exportin protein. Figure 4 shows the effect of a change in these three parameters on steady-state RISC (Figure 4 center column), pre-miRNA in the cytoplasm (Figure 4 left column), and Dicer (Figure 4 right column). The affinity of pre-miRNA towards Exportin is varied thousand fold up and down.

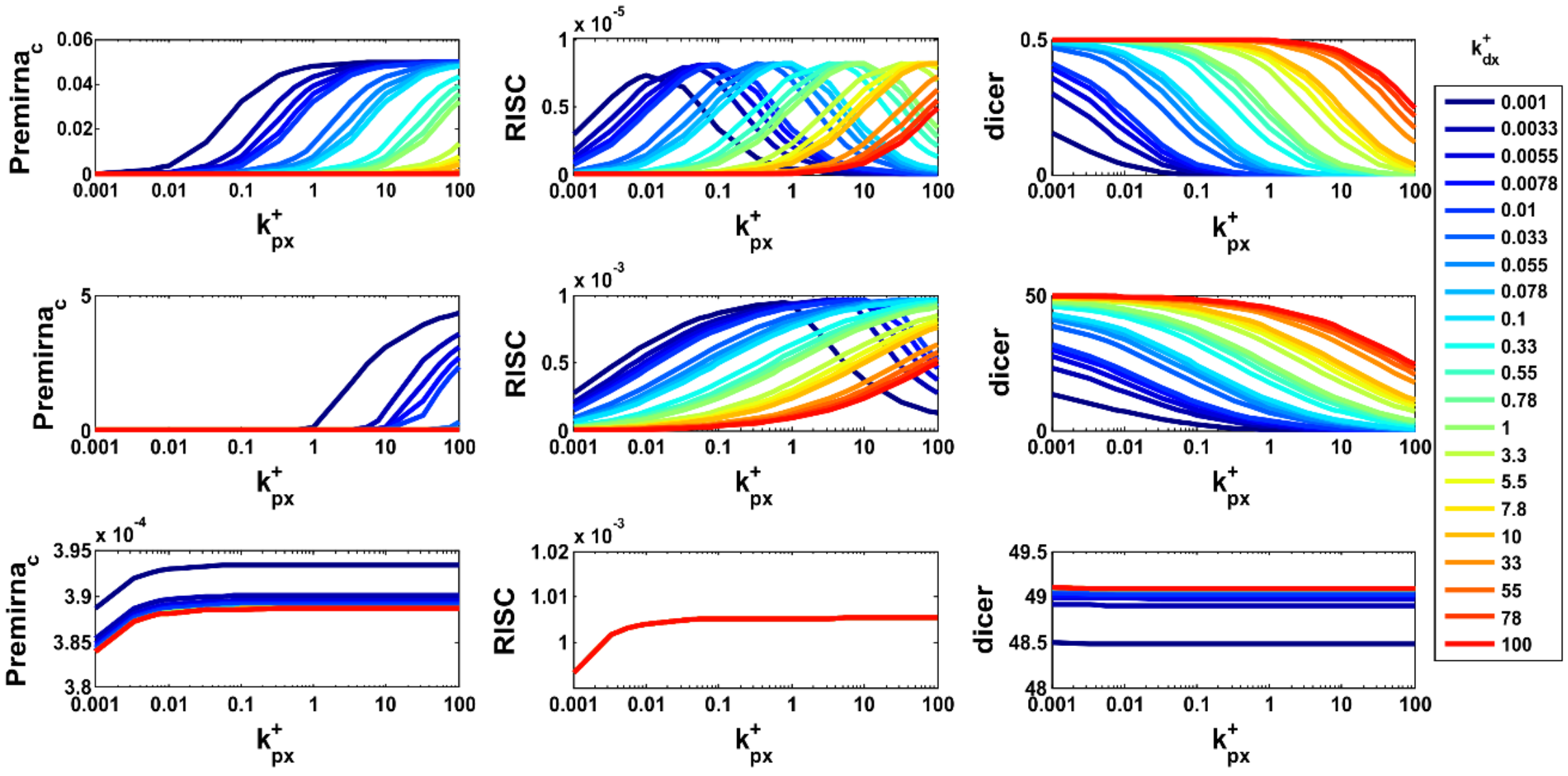

***Figure 4**: **Effect of varying pre-miRNA- Exportin and Dicer-mRNA-Exportin association rate at varied levels of Exportin- protein formation rate:** Various colored lines are for different values of Dicer:Exportin association rate. Rows from top represent increasing levels of Exportin, simulated by increasing the formation rate $b_{xpo5}$, from 100 times lower (top row) to 100-fold higher (bottom row) than is the reference value indicated in Table 3 (middle row).* denotes the point where parameters are at their reference value given in Table 3.*

As observed in the above result, at higher Exportin formation rate (Figure 4 bottom row), RISC exhibits robustness to the Dicer:Exportin association rate (Figure 4 bottom middle column: all coloured lines overlap). At lower Exportin levels (Figure 4 top panel), increasing binding affinity of pre-miRNA leads to decrease in Dicer levels and increase in pre-miRNA in the cytoplasm. RISC levels first increase and then decrease with pre-miRNA-Dicer mRNA association level, and is maximum at an optimal value of the association constant. This optimal value increases with exportin level, and increasing Dicer: Exportin affinity.

At a lower level of Exportin, increasing binding affinity of either *dicer*-mRNA or pre-miRNA leads to a reduction in the levels of the other component as both dicer-mRNA and pre-miRNA compete for Exportin protein binding. Since the formation of RISC requires both pre-miRNA and Dicer, the increase in affinity of one component constrains the second component leading to an optimum RISC level (Figure 4 middle column, top row). At higher levels of Exportin, increasing the pre-miRNA affinity causes more of the pre-miRNA to be transported. As the Exportin levels are higher, Dicer is also transported at higher levels and is present at higher concentration in the cytoplasm. As discussed, this increased Dicer level makes the pre-miRNA in the nucleus the limiting factor (Supplementary Figure 2) even at its higher transport rate. Hence, at higher exportin levels, RISC levels do not depend on the Dicer formation rate or the dicer mRNA-Xpo5 association rate.

**Positive regulation can lead to lower RISC levels**

A mechanism for positive auto-regulation of mature RISC has been identified for specific microRNAs such as *let-7* (7). The proposed mechanism is that mature RISC facilitates Argonaute-mediated enhancement of primary miRNA processing. We model this mechanism as mature RISC binding to its own primary miRNA, leading to positive auto-regulation through enhanced pri-miRNA processing. Even if this is not the actual mechanism, the mass-action kinetics for these processes captures the dependence of primary miRNA processing on mature miRNA complex concentration in the nucleus. We numerically simulated this mechanism to explore positive up-regulation for RISC. The rate of Exportin formation is assumed to be ten times more than Importin formation. Qualitatively similar results are obtained for the same Exportin and Importin formation rate as well as Importin formation rate ten times more than the exportin formation rate (data not shown).

Figure 5(A) shows the steady state RISC level relative to those for a system without any positive auto-regulation (association rate $k_{rx}^{+}$ = 0) at different levels of MP processing efficiency ($b_{prn}^{n}$).

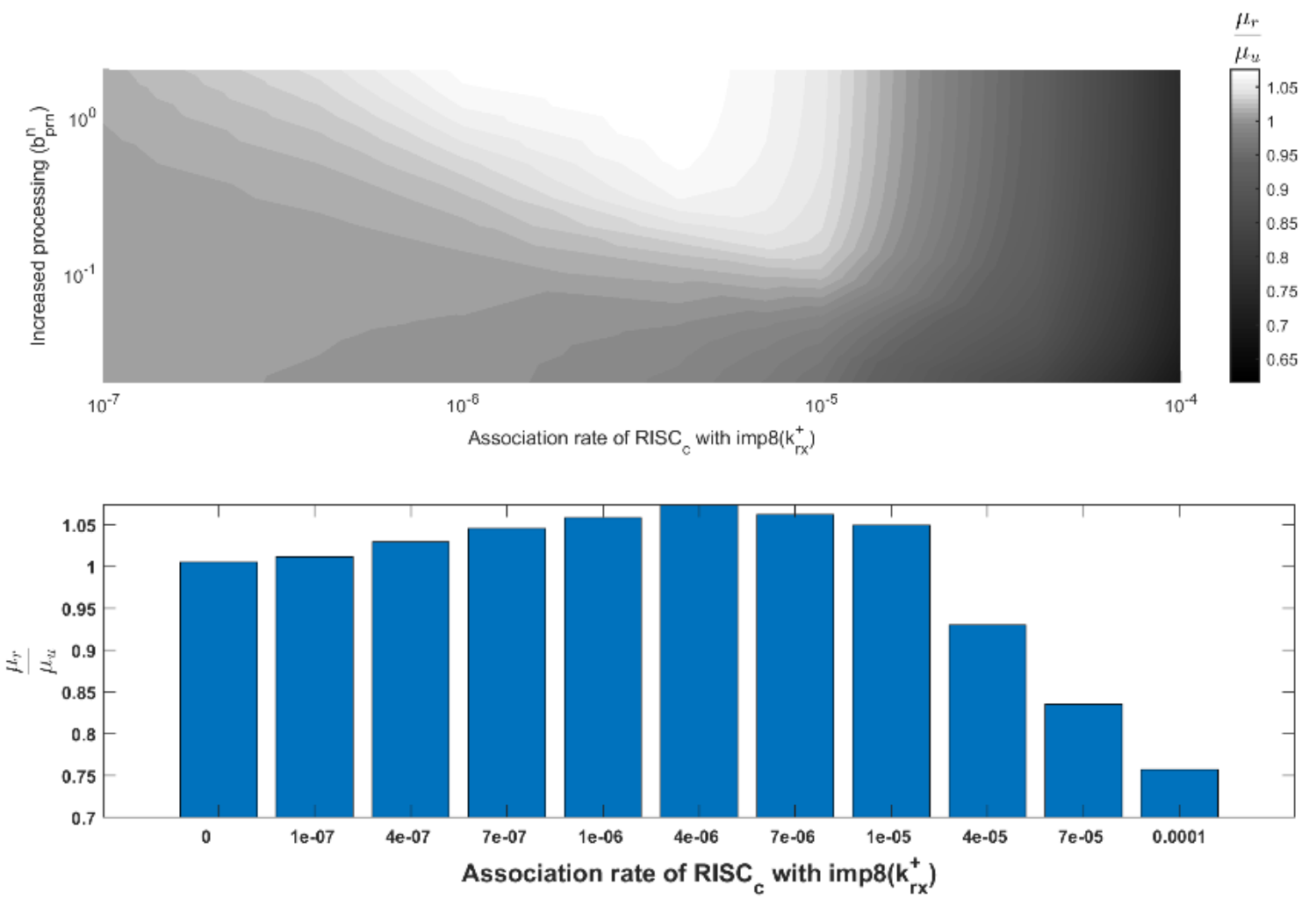

***Figure 5: Effect of enhanced processing on mature miRNA levels:*** *Steady-state levels of mature miRNA are plotted for varying mature miRNA association rate to Importin8 ($k_{rx}^{+}$) plotted along X-axis and pre-miRNA processing efficiency ($b_{prn}^{n}$) varied. along the y-axis. Increased levels of mature miRNA are observed for the optimal level of $k_{rx}^{+}$ at increased $b_{prn}^{n}$.*

Increasing auto-regulation is modeled as an increase in the association of RISC to Imp8 ($k_{rx}^{+}$) and hence the transport rate to the cytoplasm. Figure 5 (B) is the bar plot of Figure 5 (A) when MP processing efficiency is assumed to be higher ($b_{prn}^{n}$= 1 sec$^{-1}$.). We observed that at lower autoregulation efficiency, modeled as lower values of $k_{rx}^{+}$, steady-state levels of miRNA are more than the unregulated system as expected from a positive regulation. However, as the $k_{rx}^{+}$ increases, RISC decreases despite the presence of the positive auto-regulatory control. This effect is more pronounced at lower levels of $b_{prn}^{n}$.

The amount of functional RISC in the system is the free RISC present in the cytoplasm. When a large amount of RISC is transported to the nucleus for a positive regulation larger share of the RISC is in the nucleus, which can degrade and bind with high affinity to primary miRNA forming pre-miRNA. Increasing processing of pri-miRNA by MP leads to increase in RISC levels (Figure 5), however, this increase does not compensate for the RISC levels decreasing as RISC is being transported for achieving the positive auto-regulation. Hence when a large amount of miRNA is transported in the nucleus and pay off by the positively regulated loop is less, the overall amount of effective RISC is reduced. But, when a small amount of RISC is transported and pay off by the positively regulated loop is large, the overall amount of RISC is increased.

It is also observed in Figure 5 (A) that lower transport rate, when compared to a slightly higher transport rate in the positively regulated system, gives a lesser amount of RISC. Amount of RISC transported for positive feedback also plays an important role. Lower transport of RISC also cannot give maximum levels of RISC. An optimal balance of mature miRNA transport to the nucleus from cytoplasm has to be attained for the maximum amount of RISC to be formed and these simulations show the possibility that excessive transport relative to positive regulation can lead to negative regulation or a net reduction in cytoplasmic RISC. The model developed here provides a tool to computationally explore this balance between loss due to cytoplasm-to-nucleus transport, and gain due to increased Pri-miRNA processing.

**Discussion**

Simple models provide sufficiently realistic insight, but creating a comprehensive network sometimes helps to investigate network properties that cannot be observed in a simple system. We have extended a comprehensive model to explore regulated miRNA biogenesis system and the role of major components in this pathway. Many models have been formulated to explore the effect of miRNA activity, however, models explaining biogenesis of miRNA itself are few and models with regulation are even scarcer. There are many steps at which biogenesis are regulated by its own pathway components as well as other regulators. We have limited this model to include regulation only by components that are part of the biogenesis network. However there are papers that demonstrate the role of other components on the biogenesis; for instance, Siomi et al demonstrated that KH-type splicing regulatory protein(KSRP) promotes Drosha and Dicer processing (14); and Yang et al show that nuclear receptors regulated miRNA expression by interacting with the components of the miRNA biogenesis pathway either at transcriptional or posttranscriptional levels (37). Including these effects will definitely extend the scope of the model.

We simulated the mathematical model using primarily previously-defined parameters to explore the effect of varying formation rate (over-expression) of pri-miRNA as well as *dicer*-mRNA on the RISC levels. Our results were in agreement with the previously established result of Bennasser et al which demonstrated that increasing pre-miRNA leads to decrease in the Dicer level. The Bennasser et al results were not used in model construction or parameter estimation. We further analyzed the effect of Exportin on RISC levels. At lower concentration of substrate (Dicer mRNA or pre-miRNA), in general, over-expression of one of the competing substrates and the resultantly increased concentration as compared to the other competitor for transporter (Exportin) leads to competition and over-expressed competitor displaces the other component leading to increased self-transport at the expense of the other substrate. However, miRNA biogenesis is not a simple system with two substrates and one transporter. Here, one of the competitors (*dicer*-mRNA) behaves as an enzyme that catalyzes the other competitor (pre-miRNA) after its transport and translation. At lower levels of Exportin, as the non-enzymatic (nuclear pre-miRNA) competitor transport increases, some amount of the enzymatic (*dicer*-mRNA) competitor is also transported that, after being translated into Dicer, catalyzes the processing of the other competitor (cytoplasmic pre-miRNA) to give the product. Increasing levels of an enzymatic competitor expression cause a decrease in the level of the product (duplex) due to decreased availability of the non-enzymatic component. At higher levels of Exportin, whatever pre-miRNA is transported into the cytoplasm gets catalyzed by the enzymatic component to give the product (duplex). This causes pre-miRNA to become limiting and RISC becomes robust to Dicer at higher Exportin levels, as both Exportin and Dicer levels become high. These results lead to the testable predictions that (i) the Bennasser et al results will not be observed at higher Exportin levels; and (ii) At low exportin levels, increase in pre-miRNA may lead to lower RISC levels due to Dicer limitation.

We extended the analysis to study the effect of the increased binding affinity of one component (pre-miRNA) for its transport on the other component and subsequently on RISC levels. At lower levels of Exportin, increasing affinity of binding between Exportin and either component (pre-miRNA /*dicer*-mRNA) leads to an increase in RISC levels till a certain optimal value of the affinity, after which increase in affinity for either of the components leads to decrease in the RISC levels. As Dicer is required for the catalysis of pre-miRNA, increasing the level of one leads to a decrease in levels of other and vice-versa thereby affecting RISC levels. An optimal balance between both *dicer*-mRNA and pre-miRNA levels has to be maintained to achieve maximum levels of RISC since both are required for the formation of RISC. At higher concentration of Exportin, increasing the affinity of the pre-miRNA level causes more transport of pre-miRNA in the cytoplasm which is converted by enzymatic Dicer. This causes the system to operate with pseudo-first order rate and hence results in unaltered levels of RISC at steady state. This manifests as the robustness of RISC levels to changes in the affinity to Exportin, and the requirement for an optimal balance between pre-miRNA and dicer-mRNA diminishes.

We also studied the positive regulation of mature miRNA as demonstrated by Zisoulis et al. They proposed that mature miRNA is imported in nucleus and bind to its own pri-miRNA leading to its auto-regulation. Experimentally observed positive regulation can be observed in conditions where transport of mature miRNA and binding to pri-miRNA is small compared to the enhancement of the pri-mRNA processing rate due to this autoregulation. We could also identify certain conditions in which this positive loop can have a net negative effect leading to a reduction of RISC levels, for instance when the transport of RISC from the cytoplasm to nucleus is not compensated by increased pre-miRNA processing. We believe that such models that incorporate details of the biogenesis mechanism with regulatory effects will serve to further our understanding of the complex miRNA biogenesis process. Such models, combined with miRNA effect models, may lead to the discovery of non-intuitive effects resulting from the convolution of complex biogenesis dynamics and nonlinear miRNA effects.

**Materials and Methods:**

As the system contains 24 equations and is non-linear, analytical solution for the dynamics or the steady-state concentrations could not be found. The differential equations listed in Table 2, arising from a mass balance for each of the constituents of the positively regulated miRNA biogenesis system, are integrated numerically till a large time, and the resulting values of the variables used as initial guess to numerical compute the steady state values. We have performed numerical simulations using MATLAB (MathWorks, version R2012a), using functions *ode15s* for the integration and *fsolve* for the steady state calculation.

The parameter values for numerical simulation listed in Table 3 are mostly taken from Shimoni et al (36), or related models, or assumed based on the reported values for similar processes. Our conclusions from this study are independent of the absolute values of the parameters used for simulation. Specific parameter ranges used for obtaining Figures 3-5 are listed below.

For Figure 3, steady-state levels of RISC are numerically obtained by varying pri-miRNA formation ($b_{pri}$), dicer-mRNA formation ($b_{dic}^{n}$) at three different levels of Exportin protein. Three different levels of Exportin formation are thousand-fold up (0.0001 molecules $s^{-1}$) and thousand-fold down (1molecule $s^{-1}$) of the reference parameter (0.01 molecules $s^{-1}$). $b_{pri}$, $b_{dic}^{n}$ are varied five-fold up and down from the reference parameter of their formation (0.02 $s^{-1}$) rate. All the other parameters are as mentioned in Table 3. The positive regime of miRNA is simulated by keeping RISC to Imp8 association rate $k_{rx}^{+}$ = $1e^{-5}$molecules$^{-1}$ $s^{-1}$, negative regime is simulated at $k_{rx}^{+}$ =1 molecules$^{-1}$ $s^{-1}$ and unregulated $k_{rx}^{+}$ =0. The figure depicts RISC steady state when the RISC is at the positively simulated regime.

For Figure 4, RISC steady state is obtained as a function of three different parameters. The parameters varied are pre-miRNA association rate to Exportin ($k_{px}^{+}$), dicer-mRNA association rate to Exportin ($k_{dx}^{+}$) and formation rate of Exportin. Three different levels of Exportin formation are thousand-fold up (0.0001 molecules $s^{-1}$) and thousand-fold down (1 molecules $s^{-1}$) of the reference parameter (0.01 molecules $s^{-1}$). $k_{px}^{+}$ , $k_{dx}^{+}$ are varied thousand fold down and thousand-fold up from the reference parameter (1 molecules$^{-1}$ $s^{-1}$). Other parameters are as mentioned in Table 3. The positive regime of miRNA is simulated by keeping RISC to Imp8 association rate ($k_{rx}^{+}$) =1e$^{-5}$molecules$^{-1}$ s$^{-1}$, negative regime is simulated at $k_{rx}^{+}$ =1 molecules$^{-1}$s$^{-1}$ and unregulated $k_{rx}^{+}$ =0. The figure depicts RISC steady state when the RISC is at the positively simulated regime.

For Figure 5, RISC steady state is obtained by varying the association rate of RISC to Importin protein $k_{rx}^{+}$ and MP processing efficiency for RISC bound to pri-miRNA transcript ($b_{prn}^{n}$). The rate of Exportin formation is considered to be ten times more than the Importin formation rate. $k_{rx}^{+}$ is varied from 1e$^{-7}$ to thousand-fold down. MP processing efficiency is varied in the range (0.02 s$^{-1}$ - 2 s$^{-1}$), i.e. always higher than the efficiency for unbound pri-miRNA transcript. The RISC steady state values obtained are normalized to the RISC steady state value obtained during unregulated condition. A bar graph is plotted at $b_{prn}^{n}$ =1 s$^{-1}$.

**Author contributions**

DAN and CJG contributed to problem definition, formulation of the mathematical model, interpretation of results and manuscript preparation. DAN carried out the simulations and wrote the initial draft of the paper.

**Acknowledgement**

This work was supported in part by University Grants Commission (Senior Research Fellowship to DAN), DST-SERB (grant EMR/2017/003271 to CJG) and CSIR (NCL IGIB JRI grant to CJG).

**Conflict of interest disclosure**

Authors do not have any conflict of interest.

***Figure Legends:***

***Figure 1**: **Biogenesis of microRNA**. Transcription by RNA polymerase give rise to primary miRNA which is further processed to form precursor miRNA with the help of MP in the nucleus. Precursor miRNA formed is exported to the cytoplasm and following a series of step, along with the help of different enzymes, mature miRNA is generated.*

***Figure 2: Schematic representation of regulated model of miRNA biogenesis**. $RISC_C$ is the mature RISC. Red lines (⊥) indicate negative regulation. Species in blue and red colour are present in nucleus and cytoplasm respectively. Exportin and Importin molecules are assumed to be at the nuclear membrane pore. ∅ denotes Source or Sink.*

***Figure 3: Steady state levels of RISC, pri-miRNA and Dicer at varying levels of pri-miRNA ($b^n_{pri}$) and dicer-mRNA ($b^n_{dic}$) formation.** Varying colored lines are different $b^n_{dic}$. Rows from top represent increasing levels of Exportin $b_{xpo5}$. Middle row has Exportin formation rate ($b_{xpo5}$ = 0.01 molecules/s) which is the reference value indicated in Table 3. Top row is 100 times lesser and bottom row is 100 times higher than the reference value indicated in Table 3 for $b_{xpo5}$ value. *denotes RISC, pri-miRNA and Dicer-mRNA at parameter set given in Table 3. B) Reprint of Figure 4b from Bennasser et al 2011, licence agreement pending representing effect of varied small RNA concentration on dicer expression levels*

***Figure 4**: **Effect of varying pre-miRNA- Exportin and Dicer-mRNA-Exportin association rate at varied levels of Exportin- protein formation rate:** Various colored lines are for different values of Dicer:Exportin association rate. Rows from top represent increasing levels of Exportin, simulated by increasing the formation rate $b_{xpo5}$, from 100 times lower (top row) to 100-fold higher (bottom row) than is the reference value indicated in Table 3 (middle row).* denotes the point where parameters are at their reference value given in Table 3.*

***Figure 5: Effect of enhanced processing on mature miRNA levels:** Steady-state levels of mature miRNA are plotted for varying mature miRNA association rate to Importin8 ($k^+_{rx}$) plotted along X-axis and pre-miRNA processing efficiency ($b^n_{prn}$) varied. along the y-axis. Increased levels of mature miRNA are observed for the optimal level of $k^+_{rx}$ at increased $b^n_{prn}$.*

*Table 1 Comparison of the constructed model with available miRNA biogenesis models*

| **Process** | **Wang et al model** | **Barad et al model** | **This work** |
|---|---|---|---|
| Pri-miRNA formation | √ | √ | √ |
| Pri-miRNA-MP binding | Reversible | Irreversible | Reversible |
| Pri-mpcomplex degradation | Not Considered | Not Considered | √ |
| Pri-miRNA->pool | √ | √ | √ |
| *Dicer*-mRNA formation | Not Considered | Not Considered | √ |
| Exportin mediated transfer of pre-miRNA and *Dicer*-mRNA to the cytoplasm | √(transport directly) | Not Considered | Catalytic transporter model |
| Dicer mRNA and pre-miRN competition for Xpo-5 | Not Considered | Not Considered | √ |
| Nuclear pre-miRNA and Cytoplasmic premiRNA Degradation | √ | Not Considered | √ |
| Translation of Dicer-mRNA and degradation | Not Considered | Not Considered | √ |
| Pre-miRNA to duplex conversion | √(Pre-miRNA is diced to duplex) | Not Considered | Dicer-catalyzed conversion |
| Duplex degradation | √ | Not Considered | √ |
| Argonaute formation | Not Considered | Not Considered | √ |
| Argonaute degradation | Not Considered | Not Considered | √ |
| RISC formation | Reversible | Not Considered | Reversible |
| RISC degradation | √ | Not Considered | √ |
| RISC transport to nucleus and association with pri-miRNA | Not Considered | Not Considered | Reversible |
| Drosha mRNA formation and degradation | Not Considered | Not Considered | √ |
| DGCR8 mRNA formation and degradation | Not Considered | Not Considered | √ |
| Microprocessor formation | Not Considered | Not Considered | Reversible |
| Microprocessor degradation | Not Considered | Not Considered | √ |
| Exportin and Importin formation and degradation | Not Considered | Not Considered | √ |
| RISC loading of miRNA | √ | Not Considered | Not Considered |
| Effect of miRNA on mRNA | √ | Not Considered | Not Considered |
| Microprocessor off-target effects | Not Considered | √ | Not Considered |

*Table 2 Ordinary differential equation for the regulated biogenesis system. Terms in bold are non-zero in the regulated system*

| |
|---|
| $\frac{pri}{dt} = b_{pri} - k^{+}_{pm} pri * mp + k^{-}_{pm} PMC - d_{pri} pri \mathbf{- k^{+}_{pr} risc_{n} * pri + k^{-}_{pr} PRC}$ |
| $\frac{pre_{n}}{dt} = b^{n}_{pre} PMC - k^{+}_{px} pre_{n} * xpo5 + k^{-}_{px} prexpocomp - d^{n}_{pre} pre_{n} \mathbf{+ b^{n}_{prm} RMP}$ |
| $\frac{pre_{c}}{dt} = b^{c}_{pri} PXC * \frac{V_{n}}{V_{c}} - d^{c}_{pre} pre_{c} - k^{+}_{pd} dicer * pre_{c} + k^{-}_{pd} PDC$ |
| $\frac{PXC}{dt} = k^{+}_{px} pre_{n} * xpo5 - k^{-}_{px} PXC - d_{px} PXC - b^{c}_{pri} PXC * \frac{V_{n}}{V_{c}}$ |
| $\frac{mDicer_{n}}{dt} = b^{n}_{dic} - k^{+}_{dx} mDicer_{n} * xpo5 + k^{-}_{dx} DXC - d^{n}_{dic} mDicer_{n}$ |
| $\frac{mDicer_{c}}{dt} = b^{c}_{dic} DXC * \frac{V_{n}}{V_{c}} - d^{c}_{dic} mDicer_{c}$ |
| $\frac{dicer}{dt} = b_{dic} mDicer_{c} - d_{dic} dicer - k^{+}_{pd} dicer * pre_{c} + k^{-}_{pd} PDC + b_{du} PDC$ |
| $\frac{PDC}{dt} = k^{+}_{pd} dicer * pre_{c} - k^{-}_{pd} PDC - d_{pd} PDC - b_{du} PDC$ |
| $\frac{xpo5}{dt} = -k^{+}_{px} pre_{n} * xpo5 + k^{-}_{px} PXC + b^{c}_{pri} PXC * \frac{V_{n}}{V_{c}} - k^{+}_{dx} mDicer_{n} * xpo5 + k^{-}_{px} DXC + b^{c}_{dic} DXC + b_{xpo5} - d_{xpo5} xpo5$ |
| $\mathbf{\frac{imp8}{dt} = -k^{+}_{rx} risc_{c} * imp8 + k^{-}_{rx} RIC + b^{n}_{r} RIC \frac{V_{c}}{V_{n}} + b_{imp8} - d_{imp8} imp8}$ |
| $\frac{duplex}{dt} = b_{du} PDC - d_{du} duplex - k^{+} duplex * ago + k^{-} risc_{c}$ |
| $\frac{ago}{dt} = b_{ago} - d_{ago} ago - k^{+} duplex * ago + k^{-} risc_{c}$ |
| $\frac{risc_{c}}{dt} = k^{+} duplex * ago - k^{-} risc_{c} - d^{c}_{r} risc_{c} \mathbf{- k^{+}_{rx} risc_{c} * imp8 + k^{-}_{rx} RIC}$ |
| $\frac{risc_n}{dt} = b_r^n RIC \frac{V_c}{V_n} - d_r^n risc_n - k_{pr}^+ risc_n * pri + k_{pr}^- PRC + b_{prn}^n RMP$ |
| $\frac{DXC}{dt} = k_{dx}^+ mDicer_n * xpo5 - k_{dx}^- DXC - d_{dx} DXC - b_{dic}^c DXC * \frac{V_n}{V_c}$ |
| $\frac{mDrosha}{dt} = d_{drm} - d_{drm} mDrosha$ |
| $\frac{mDGCR8}{dt} = \frac{b_{dgm}}{1 + \frac{mp}{kin3}} - d_{dgm} mDGCR8$ |
| $\frac{Drosha}{dt} = b_{drp} mDrosha - \frac{b_{drp} Drosha}{1 + \frac{DGCR8}{kin2}} - k_{mp}^+ Drosha * DGCR8 + k_{mp}^- mp$ |
| $\frac{DGCR8}{dt} = b_{dgp} mDGCR8 - b_{dgp} DGCR8 - k_{mp}^+ Drosha * DGCR8 + k_{mp}^- mp$ |
| $\frac{mp}{dt} = k_{pm}^- PMC - k_{mp}^+ pri * mp + b_{pre}^n PMC + k_{mp}^+ Drosha * DGCR8 - k_{mp}^- mp - d_{mp} mp - k_{prm}^+ PRC * mp + k_{prm}^- RMP + b_{prm}^n RMP$ |
| $\frac{RIC}{dt} = k_{rx}^+ risc_c * imp8 - k_{rx}^- RIC - d_{rx} RIC - b_r^n RIC \frac{V_c}{V_n}$ |
| $\frac{PRC}{dt} = k_{pr}^+ risc_n * pri - k_{pr}^- PRC - d_{pr} PRC - k_{prm}^+ PRC * mp + k_{prm}^- RMP$ |
| $\frac{PMC}{dt} = k_{mp}^+ pri * mp - k_{pm}^- PMC - d_{pm} PMC - b_{pre}^n PMC$ |
| $\frac{RMP}{dt} = k_{prm}^+ PRC * mp - k_{prm}^- RMP - d_{prm} RMP - b_{prm}^n RMP$ |

*Table 3 Reaction rate parameters used for numerical simulations.*

| S. No | Reaction | Parameter | Value | Parameter/Process in Shimoni's model |
|---|---|---|---|---|
| 1 | pool →primiRNA | $b_{pri}$ | 0.02 molecules $s^{-1}$ | mRNA transcription |
| 2 | primiRNA + MP → pri-mp-complex | $k_{pm}^{+}$ | 1 molecules $^{-1}s^{-1}$ | complex association |
| 3 | PMC →primiRNA + MP | $k_{pm}^{-}$ | 0.02 $s^{-1}$ | complex dissociation |
| 4 | PMC→ pool | $d_{pm}$ | 0.002 $s^{-1}$ | mRNA degradation |
| 5 | PMC→ $pre_n$ +mp | $b_{pre}^{n}$ | 0.02 $s^{-1}$ | complex dissociation |
| 6 | primiRNA → pool | $d_{pri}$ | 0.002 $s^{-1}$ | mRNA degradation |
| 7 | $pre_n$ +xpo5 → PXC | $k_{px}^{+}$ | 1 molecules$^{-1}s^{-1}$ | complex association |
| 8 | PXC → $pre_n$ +xpo5 | $k_{px}^{-}$ | 0.02 $s^{-1}$ | complex dissociation |
| 9 | PXC → pool | $d_{px}$ | 0.002 $s^{-1}$ | mRNA degradation |
| 10 | PXC → $pre_c$ +xpo5 | $b_{pri}^{c}$ | 0.02 $s^{-1}$ | complex dissociation |
| 11 | $pre_c$ → pool | $d_{pre}^{c}$ | 0.002 $s^{-1}$ | mRNA degradation |
| 12 | $pre_n$ → pool | $d_{pre}^{n}$ | 0.002 $s^{-1}$ | mRNA degradation |
| 13 | pool → $mDicer_n$ | $b_{dic}^{n}$ | 0.02 moleculess$^{-1}$ | mRNA transcription |
| 14 | $mDicer_n$ +xpo5→DXC | $k_{dx}^{+}$ | 1 molecules$^{-1}s^{-1}$ | complex association |
| 15 | DXC→ $mDicer_n$ +xpo5 | $k_{dx}^{-}$ | 0.02 $s^{-1}$ | complex dissociation |
| 16 | DXC→ pool | $d_{dx}$ | 0.002 $s^{-1}$ | mRNA degradation |
| 17 | DXC→ $mDicer_c$ +xpo5 | $b_{dic}^{c}$ | 0.02 $s^{-1}$ | complex dissociation |
| 18 | $mDicer_c$ → Dicer | $b_{dic}$ | 0.01 $s^{-1}$ | mRNA translation |
| 19 | $mDicer_n$ → pool | $d_{dic}^{n}$ | 0.002 $s^{-1}$ | mRNA degradation |
| 20 | $mDicer_c$ → pool | $d_{dic}^{c}$ | 0.002 $s^{-1}$ | mRNA degradation |
| 21 | Dicer→ pool | $d_{dic}$ | 0.001 $s^{-1}$ | Protein degradation |
| 22 | Dicer+ $pre_c$ → PDC | $k_{pd}^{+}$ | 1 molecules$^{-1}s^{-1}$ | complex association |
| 23 | PDC→ $pre_c$ +Dicer | $k_{pd}^{-}$ | 0.02 $s^{-1}$ | complex dissociation |
| 24 | PDC→ pool | $d_{pd}$ | 0.002 $s^{-1}$ | mRNA degradation |
| 25 | PDC →duplex+ Dicer | $d_{du}$ | 0.02 $s^{-1}$ | complex dissociation |
| 26 | duplex→ pool | $d_{du}$ | 0.002 $s^{-1}$ | mRNA degradation |
| 27 | pool →Ago | $b_{ago}$ | 0.01 molecules $s^{-1}$ | mRNA translation |
| 28 | Ago→ pool | $d_{ago}$ | 0.001 $s^{-1}$ | Protein degradation |
| 29 | Duplex + Ago → $risc_c$ | $k^{+}$ | 1 molecules$^{-1}s^{-1}$ | complex association |
| 30 | $risc_c$ → duplex + Ago | $k^{-}$ | 0.01 $s^{-1}$ | complex dissociation |
| 31 | $risc_c$ → pool | $d_{r}^{c}$ | 0.002 $s^{-1}$ | mRNA degradation |
| 32 | $risc_c$ +imp8→ RIC | $k_{rx}^{+}$ | 1 molecules$^{-1}s^{-1}$ | complex association |
| 33 | RIC → $risc_c$ +imp8 | $k^{-}_{rx}$ | 0.02 $s^{-1}$ | complex dissociation |
| 34 | RIC→Pool | $d_{rx}$ | 0.002 $s^{-1}$ | mRNA degradation |
| 35 | RIC → $risc_n$ +imp8 | $b^{n}_{r}$ | 0.02 $s^{-1}$ | complex dissociation |
| 36 | $risc_n$ → Pool | $d^{n}_{r}$ | 0.002 $s^{-1}$ | Protein degradation |
| 35 | pool → drosha mRNA | $b_{drm}$ | 0.02molecules $s^{-1}$ | mRNA transcription |
| 36 | drosha mRNA → Drosha | $b_{drp}$ | 0.01 $s^{-1}$ | mRNA translation |
| 37 | drosha mRNA→ pool | $d_{drm}$ | 0.002 $s^{-1}$ | mRNA degradation |
| 38 | Drosha→pool | $d_{drp}$ | 0.001 $s^{-1}$ | Protein degradation |
| 39 | pool → DGCR8 mRNA | $b_{dgm}$ | 0.02molecules $s^{-1}$ | mRNA transcription |
| 40 | DGCR8 mRNA→ DGCR8 | $b_{dgp}$ | 0.01 $s^{-1}$ | mRNA translation |
| 41 | DGCR8 mRNA→ pool | $d_{dgm}$ | 0.002 $s^{-1}$ | mRNA degradation |
| 42 | DGCR8→ pool | $d_{dgp}$ | 0.001 $s^{-1}$ | Protein degradation |
| 43 | DGCR8 + Drosha→ MP | $k^{+}_{mp}$ | 1 $molecules^{-1}s^{-1}$ | complex association |
| 44 | MP →DGCR8 + Drosha | $k^{-}_{mp}$ | 0.02 $s^{-1}$ | complex dissociation |
| 45 | MP →pool | $d_{mp}$ | 0.001 $s^{-1}$ | Protein degradation |
| 46 | pool → xpo5 | $b_{xpo5}$ | 0.01 molecules $s^{-1}$ | mRNA translation |
| 47 | xpo5 →pool | $d_{xpo5}$ | 0.001 $s^{-1}$ | Protein degradation |
| 48 | pool →imp8 | $b_{imp8}$ | 0.01molecules $s^{-1}$ | mRNA translation |
| 49 | imp8→ pool | $d_{imp8}$ | 0.001 $s^{-1}$ | Protein degradation |
| 50 | $risc_n$ +primiRNA → PRC | $k^{+}_{pr}$ | 1000 $molecules^{-1}s^{-1}$ | Assumed |
| 51 | PRC → $risc_n$ +primiRNA | $k^{-}_{pr}$ | 0.02 $s^{-1}$ | complex dissociation |
| 52 | PRC →pool | $d_{pr}$ | 0.002 $s^{-1}$ | mRNA degradation |
| 53 | PRC+MP→ RMP | $k^{+}_{prm}$ | 1 $molecules^{-1}s^{-1}$ | complex association |
| 54 | RMP → PRC+MP | $k^{-}_{prm}$ | 0.02 $s^{-1}$ | complex dissociation |
| 55 | RMP → Pool | $d_{prm}$ | 0.002 $s^{-1}$ | mRNA degradation |
| 56 | RMP → $pre_n$ + MP + $risc_n$ | $b^{n}_{prn}$ | 0.02-2 $s^{-1}$ | Assumed (greater than $b^{n}_{pre}$ ) |
| 59 | volume of the nucleus ($L^1$) | | $4X10^{-13}$ | |
| 60 | volume of cytoplasm($L^1$) | | $4X10^{-12}$ | |

**Note:** *Parameter No 50 and 56 are assumed and Nos 59-60 taken from Wang et al. (35). All other parameters are based on the values used by Shimoni et al (36) for analogous processes. The numerical value is identical and the process for which it is used in (36) is indicated in the last column.*

# Supplementary Figures

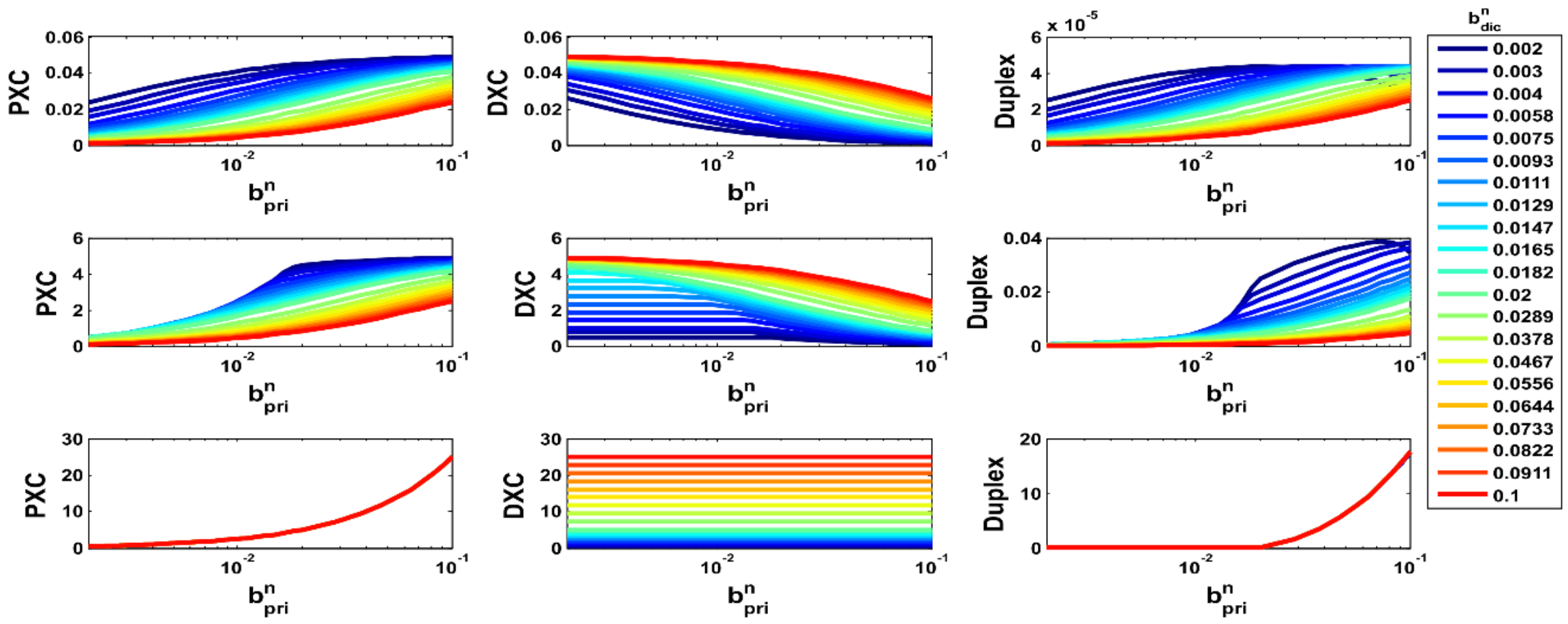

***Supplementary Figure 1: Steady state levels of PremiRNA:exportin complex (PXC), DicermRNA:exportin complex (DXC) and Duplex at varying levels of pri-miRNA ($b_{pri}$) and dicer-mRNA ($b_{dic}$) formation.*** *Varying colored lines are different $b_{dic}$ . Rows from top represent increasing levels of Exportin ($b_{xpo5}$) . Middle row has Exportin formation rate $b_{xpo5}$ = 0.01 molecules/s which is the reference value indicated in Table 3. Top row is 100 times lesser and bottom row is 100 times higher than the reference value indicated in Table 3 for the $b_{xpo5}$ value.*

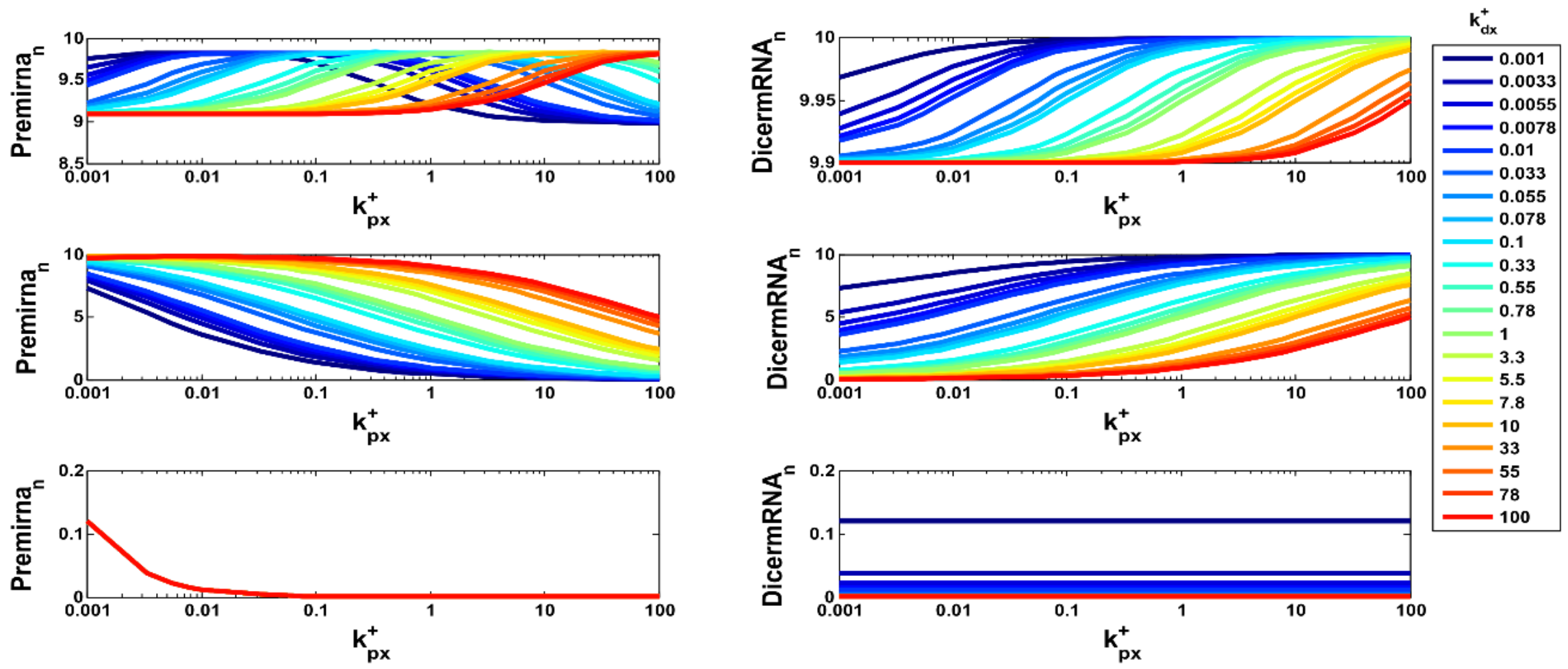

***Supplementary Figure 2**: **Effect of varying pre-miRNA- Exportin and Dicer-mRNA-Exportin association rate at varied levels of Exportin- protein formation rate on PremiRNA$_n$ and DicermRNA$_n$:*** *Various colored lines are for different values of Dicer:Exportin association rate. Rows from top represent increasing levels of Exportin, simulated by increasing the formation rate $b_{xpo5}$, from 100 times lower (top row) to 100-fold higher (bottom row) than is the reference value indicated in Table 3 (middle row)*